\documentclass[myepj-spec]{mySvjour}
\usepackage{graphicx}
\usepackage{hyperref}
\usepackage{color}
\usepackage{xspace}
\usepackage[square,sort&compress,numbers]{natbib}   
\usepackage{verbatim}
\usepackage{amsmath,amssymb,amsfonts} 
\usepackage{tabularx}
\usepackage{multicol}
\usepackage{footnote}
\usepackage{subcaption}
\usepackage{booktabs}
\usepackage{listings}
\usepackage{graphicx} 
\usepackage{xcolor}
\usepackage{tikz}
\usepackage{siunitx}
\usepackage{physics}
\AtBeginDocument{\RenewCommandCopy\qty\SI}
\ExplSyntaxOn
\msg_redirect_name:nnn { siunitx } { physics-pkg } { none }
\ExplSyntaxOff
\usetikzlibrary{positioning,arrows.meta}
\tikzset{
  box/.style   = { rectangle, rounded corners, draw},
  valve/.style = {-{Triangle[fill=white,width=1em,
      length=1em]Triangle[fill=white,width=1em,length=1em,reversed]}},
}
\tikzstyle{arrow} = [->,>=stealth]

\usepackage{lmodern,bm}                
\usepackage[T1]{sansmath} 
\SetMathAlphabet{\mathsfbf}{sans}{\sansmathencoding}{\sfdefault}{bx}{sl}
\usepackage{etoolbox}
\AtBeginEnvironment{sansmath}{}{}{}

\usepackage{lipsum}
\usepackage{lineno}

\newcommand{\Supax}{\textsc{Supax\,}}

\newcommand{\muev}{\ensuremath{\mu\textrm{eV}}}

\newcommand{\CDR}{\cite{SupaxCDR}}

\newcommand{\excl}{\expval{\cos^2 \theta}_T^\text{excl.}}

\definecolor{darkblue1}{rgb}{0,0,.2}
\definecolor{darkblue}{rgb}{0,0,.2}
\definecolor{darkred}{rgb}{0.5,0,0}
\pagecolor{white} 
\color{black}     
\hypersetup{breaklinks=true, 
	colorlinks=true, 
	linkcolor=darkblue1, 
	menucolor=darkblue1, 
	urlcolor=darkblue1,
	citecolor=darkblue1,
	pdftitle={},
	pdfauthor={},
	pdfsubject={},
	pdfkeywords={},
	pdfproducer={}
}

\usepackage{siunitx}

\bibstyle{plain}

\begin{document}
	
	{%
		\begin{@twocolumnfalse}
			
			\begin{flushright}
				\normalsize
			\end{flushright}
			
			\vspace{-2cm}
			\title{\Large\boldmath First Axion Search Results of the SUPAX Prototype Experiment}
\author{Tim Schneemann$^{1}$\footnote{tischnee@uni-mainz.de}, Hendrik Bekker$^{2,3,4}$, Dmitry Budker$^{2,3,4,5}$, Kristof Schmieden$^{6}$, Matthias Schott$^{6}$, Malavika Unni$^{3,4}$, Arne Wickenbrock$^{2,3,4}$ \footnote{kristof.schmieden@cern.ch}}
\institute{
\inst{1} Institute of Physics, Johannes Gutenberg University, Mainz, Germany \\
\inst{2} Excellence Cluster PRISMA++, Johannes Gutenberg-Universit{\"a}t Mainz, 55128 Mainz, Germany\\
\inst{3} Helmholtz-Institut Mainz, 55099 Mainz, Germany \\
\inst{4} GSI Helmholtzzentrum für Schwerionenforschung GmbH, 64291 Darmstadt, Germany\\
\inst{5} Department of Physics, University of California, Berkeley, California 94720, USA\\\inst{6} Rheinische Friedrich-Wilhelms-University, Bonn, Germany
}


			\abstract{The SUPerconduction AXion search experiment (\textsc{Supax}) is a future haloscope-type detector designed to probe axion-like particles (ALPs) as candidates for dark matter and solutions to the strong-CP problem in the mass range between $8\,\muev$ and 30\,$\mu$eV. In the course of the preparation of \Supax, a prototype experiment was built and operated. Using a copper cavity, cooled down to a temperature of 2\,K and operated in a magnetic field of 12\,T, we probe axion masses around $34\,\muev$ and exclude axion-photon couplings down to $\abs{g_{a\gamma\gamma}}> 1.6\cdot 10^{-13}\textrm{GeV}^{-1}$. The data was also used to  exclude dark photons in the same mass range with a kinetic mixing parameter of $\chi > 1.4\cdot 10^{-12}$. Details of the experimental setup and the analysis strategy are summarized in this paper.}	
	\maketitle
	\end{@twocolumnfalse}
}

\tableofcontents
\vspace{0.5cm}

\section{Introduction}
\label{sec:intro}

The QCD axion is a hypothetical pseudoscalar particle originally proposed to resolve the strong CP problem in quantum chromodynamics (QCD) \cite{Peccei:1977hh,Wilczek:1977pj,Weinberg:1977ma}, which questions why the strong interaction appears to be fine-tuned to guarantee CP conservation. Beyond this fundamental theoretical motivation, the axion also emerges as a compelling dark matter candidate \cite{PRESKILL1983127,ABBOTT1983133,DINE1983137,Ipser:1983mw,Turner:1983sj}. Various cosmological models predict that axions produced in the early universe could constitute a significant fraction, or even the entirety, of the observed dark matter. These models also relate the axion mass to its relic abundance, with many simulations favoring masses in the range of a few tens of $\mu$eV and above \cite{Saikawa:2024bta}.

A broad range of experimental strategies have been developed to search for axions, most of which exploit their predicted coupling to photons via the inverse Primakoff effect \cite{Pirmakoff:1951pj}. In this process, axions can convert into photons when interacting with a strong magnetic field, providing a basis for resonant detection schemes.

Among the most promising and widely pursued techniques is the haloscope approach \cite{Sikivie:1983ip}. In a haloscope, axions from the galactic dark matter halo are expected to convert into microwave photons within a resonant cavity immersed in a strong magnetic field. The cavity is tuned to match the expected axion mass, maximizing the conversion signal. Several state-of-the-art experiments, such as ADMX \cite{ADMX:2020ote,2010PhRvL.104d1301A}, HAYSTACK \cite{HAYSTAC:2018rwy,HAYSTAC:2024jch}, RADES \cite{Ahyoune:2024klt}, QUAX \cite{QUAX:2024fut}, and CAPP \cite{CAPP-2024-0,CAPP-2024-1} have successfully employed this technique and continue to probe axion masses across different frequency bands.

In this context, the SUPerconduction AXion search experiment (\textsc{Supax}) \cite{SupaxCDR} is being developed as a next-generation haloscope at the University of Bonn. As a first step, a prototype experiment has been constructed and operated to test key components and evaluate system performance. The prototype employs a 14.1\,T superconducting solenoid magnet, a precision-machined copper cavity with a quality factor of $5\cdot 10^4$ and a resonance frequency of 8.3\,GHz, and a low-noise signal readout with transistor based amplifiers. The entire setup is operated at liquid helium (LHe) temperatures to minimize thermal noise and enhance sensitivity. In this paper, we present the design and first experimental results from the \Supax prototype, laying the groundwork for the full-scale experiment.

This paper is structured as follows. After a brief explanation of the expected signal in Sec.\,\ref{sec:theory} the experimental setup is detailed in Sec,\,\ref{sec:setup}. Data analysis and results are discussed in Sec.\,\ref{sec:analysis}, followed by an outlook to the next stages of the experiment. \\

\section{Theoretical Background: Axion Signal in a Resonant Cavity}
\label{sec:theory}

Axions and axion-like particles (ALPs) are pseudoscalar fields \( a(x) \) predicted to couple weakly to photons via the Lagrangian term
\begin{equation}
\mathcal{L}_{a\gamma\gamma} = -\frac{1}{4} \abs{g_{a\gamma\gamma}} \, a \, F_{\mu\nu} \tilde{F}^{\mu\nu} = \abs{g_{a\gamma\gamma}} \, a \, \mathbf{E} \cdot \mathbf{B}\,,
\end{equation}
where \( \abs{g_{a\gamma\gamma}} \) is the axion-photon coupling constant, \( F_{\mu\nu} \) is the electromagnetic field strength tensor, \( \tilde{F}^{\mu\nu} \) is its dual, and \( \mathbf{E} \) and \( \mathbf{B} \) are the electric and magnetic fields, respectively. In the presence of a background magnetic field \( \mathbf{B}_0 \), this interaction leads to the conversion of axions into photons, which forms the basis for haloscope experiments.

Assuming the standard cosmological halo model where axions with mass \( m_a \) play the role of cold dark matter \cite{Marsh:2015xka,DiLuzio:2020wdo}, the axion field can be approximated as a classical, spatially coherent oscillating background: $a(t) = a_0 \cos(m_a t)$,
where \( a_0 \) is the amplitude, related to the local dark matter density \( \rho_a \) via
$\rho_a =  m_a^2 a_0^2/2$.

Inside a resonant microwave cavity permeated by a static magnetic field \( \mathbf{B}_0 \), the axion field acts as a source term for Maxwell’s equations, inducing an oscillating electric field at the frequency corresponding to the axion mass. The resulting power deposited into a resonant mode of the cavity is \cite{Sikivie:1985yu}
\begin{equation}
 P_{\textrm{sig}} = \abs{g_{a\gamma\gamma}}^2 \rho_a \frac{\beta}{\beta + 1} \frac{1}{m_a} B_0^2 V C Q_{\textrm{eff}} \eta \, \hbar^2 c^5 \varepsilon_0 \label{eq:PS_axions}\,,
\end{equation}
where $B_0$ is the external magnetic field, $V$ is the volume of the cavity, $C$ is the form factor of the cavity mode, defined as
    \begin{equation}
    C = \frac{\left| \int_V \mathbf{E}_c \cdot \mathbf{B}_0 \, dV \right|^2}{V B_0^2 \int_V \epsilon(\mathbf{x}) |\mathbf{E}_c|^2 dV}\,.
    \end{equation}
The form factor is a measure of the overlap of the external magnetic field with the electric field of the cavity mode $\mathbf{E}_c$, where $\epsilon(\mathbf{x})$ is the permittivity in the resonant cavity.     
$Q_\text{eff}$ is defined as $ Q_\text{eff}= \min(Q_L, Q_a)$, where $Q_L$ is the loaded quality factor of the cavity, and $Q_a \sim 10^6$ is the effective quality factor of the axion signal due to its narrow energy spread, and $\eta$ summarizes any experimental losses.

The expected signal is a narrow-band, quasi-monochromatic peak centered at a frequency \( \nu_a = \frac{m_a c^2}{h} \) with a relative linewidth \( \Delta \nu / \nu_a \sim 10^{-6} \), corresponding to the virial velocity distribution of dark matter particles in the galactic halo. The noise term is given by

\begin{equation}
k_B T_{\text{sys}} = h \nu \left( \frac{1}{e^{h \nu / k_B T} - 1} + \frac{1}{2} + N_A \right)\,,
\end{equation}
where $N_A$ is the noise from the amplifier, and T is the physical temperature of the system. The signal-to-noise ratio (SNR) for an integration time \( t \) is then calculated as
\begin{equation}
\text{SNR} = \frac{P_\text{sig}}{k_B T_\text{sys}} \sqrt{\frac{t}{\Delta \nu}}\,,\label{eq:snr}
\end{equation}
where \( T_\text{sys} \) is the system noise temperature and \( \Delta \nu \sim \nu_a / Q_a \) is the bandwidth of the axion signal.

Maximizing sensitivity thus requires a high magnetic field \( B_0 \), large cavity volume \( V \), high-quality factor \( Q_L \), and low system noise temperature \( T_\text{sys} \). The \Supax experiment is designed with these requirements in mind, targeting axion masses in the range \( 8\,\mu\text{eV} \lesssim m_a \lesssim 30\,\mu\text{eV} \) using tunable, superconducting cavities within a 12\,T solenoidal magnetic field at 10\,mK ambient temperature. The prototype experiment has been performed with $B=12.1$ T, $T_{sys}$=9.99\,K at 2\,K ambient temperature and a resonance frequency of 8.27\,GHz. \\

\newpage
\section{Experimental Setup}
\label{sec:setup}

\begin{figure}[ht]
\centering
\begin{minipage}{9.2cm}
  \centering
    \includegraphics[width=0.99\linewidth]{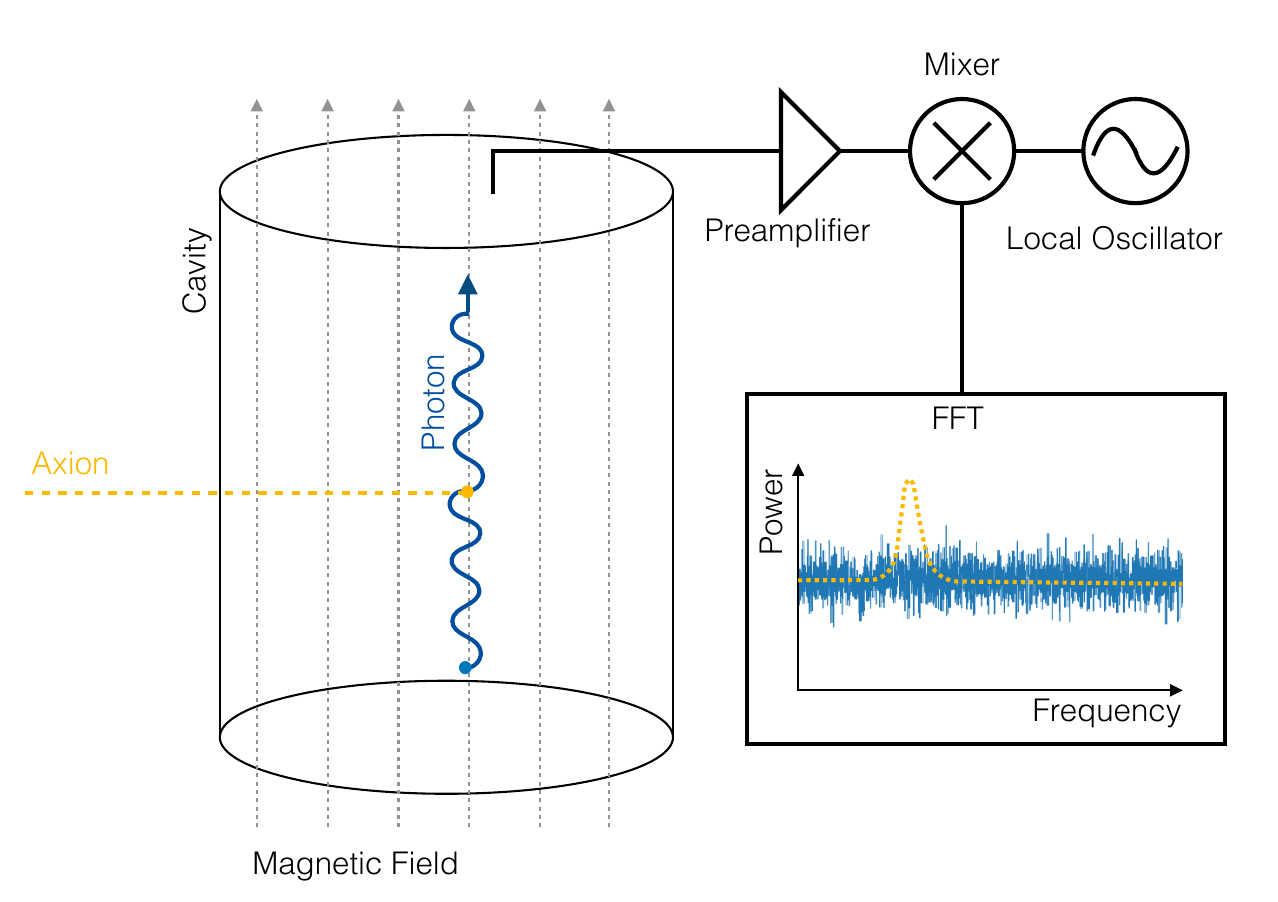}
    \caption{Illustration of the basic principle of a haloscope search experiment.}
    \label{fig:principle}

\end{minipage}%
\hspace{0.2cm}
\begin{minipage}{7.3cm}
  \centering
    \includegraphics[width=0.99\linewidth]{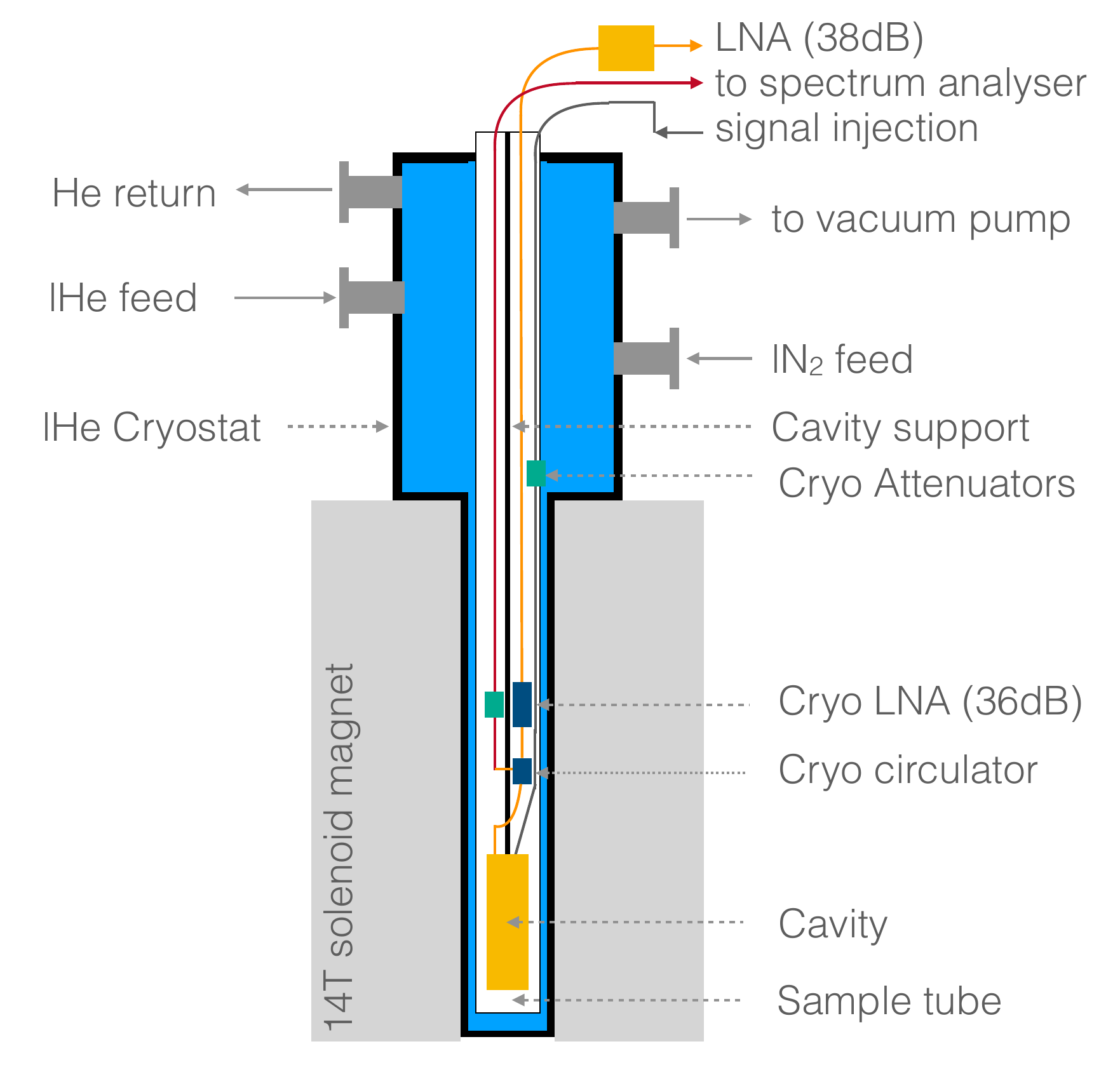}
    \caption{Scetch of the experimental setup of the prototype Experiment for \Supax.\vspace{0.5cm}}
    \label{fig:setup_scheme}
\end{minipage}
\end{figure}

The basic principle of a haloscope experiment is schematically illustrated in Fig.\,\ref{fig:principle}, where the axion conversion in a magnetic field within an electromagnetic cavity is shown. 
The resulting signal is amplified, converted to a lower-frequency band, and digitized. The resulting time-series is Fourier-transformed into the frequency domain and analysed for any excess power around the resonance frequency. 
This setup is implemented as a prototype of the \Supax experiment located at the University of Mainz; it utilises a liquid-helium cryostat inserted into the room-temperature bore of a 14.1\,T magnet located at the Helmholtz-Institute Mainz. A schematic drawing of the setup is shown in Fig.\,\ref{fig:setup_scheme}. The temperature of the cryostat can be adjusted between 1.5\,K and 300\,K as needed and is stabilised to better than 0.1\,K. The magnetic field can be adjusted between 0 and 14.1\,T with a ramp time of approximately 5\,h.

The cavity has outer dimensions of $\approx160 \times 40 \times \SI{26.8}{mm^3}$ optimised to fully utilize the available space while keeping the $\text{TM}_{010}$ mode separated from the next mode by at least $\SI{100}{MHz}$ \cite{SupaxCDR}. This leads to a usable volume of $150 \times 22.8 \times 30\, \si{mm^3}$ where the corners are rounded ($r = \SI{9}{mm}$) to reduce boundary effects. The resulting resonance frequency of the $\text{TM}_{010}$ mode is $f = \SI{8.27}{GHz}$ at cryogenic conditions,  corresponding to an axion mass of $m_{A'} = 2 \pi \hbar \cdot f \approx \SI{34}{\mu eV}$.

During cooling of the cavity to $T = \SI{2}{K}$ it is filled with helium vapour at an adjustable pressure. The resonance frequency shifts from $\SI{8.4}{GHz}$ to $\SI{8.27}{GHz}$ as the effect of the dielectric property of liquid helium vapour compensates for the frequency increase due to thermal contraction of the cavity itself. This effect allows tuning of the resonance frequency via changes in helium pressure, as described in \cite{SupaxCDR}.

The cavity has one critically and one weakly coupled port. 
The critically coupled antenna ($\beta = 1$) is connected via a circulator to a cryogenic low-noise amplifier (LNF-LNC4\_16B) with \SI{36}{dB} gain, followed by a real-time spectrum analyser (RSA) with an internal SI{25}{dB} amplifier outside the cryostat. For characterizing the cavity, a network analyser is connected to both cavity ports as well as the third port of the circulator, allowing a full set of S--parameter measurements. In this configuration the weakly coupled antenna ($\beta \ll 1$) is used to inject signals from a Vector Network Analyser (VNA) via a \SI{20}{dB} attenuator.
The measured S--parameters agree well with values obtained by finite-element simulation, which is also used to optimise the antenna lengths. 
The unloaded quality factor is determined to be $Q_0 = 5.2(2)\times 10^{4}$ and the coupling $\beta$ of the readout antenna is measured to be $\beta = 0.87 \pm 0.06$.

The RSA is connected via a USB3 link to a dedicated readout personal computer (PC) and is controlled via custom software. 
A 10\,MHz window around the centre frequency is read out by the RSA and sampled at  a rate of 28\,MS/s. 
The real-time data of both quadratures (IQ) are streamed from the device and converted via a fast Fourier transform into the frequency domain with a readout bandwidth of 1\,kHz. The resulting spectra from 1\,s intervals are averaged in real time and stored for offline analysis. 

\section{Data Analysis and Results}
\label{sec:analysis}

The details of the analysis methodology, as well as the data format and specifics of the data acquisition are discussed in the conceptual design report \CDR. The experimental parameters relevant for the axion data taking are summarized in Tab. \ref{tab:experimental_parameters}.

\begin{figure}[ht]
    \centering
    \includegraphics[width=0.6\linewidth]{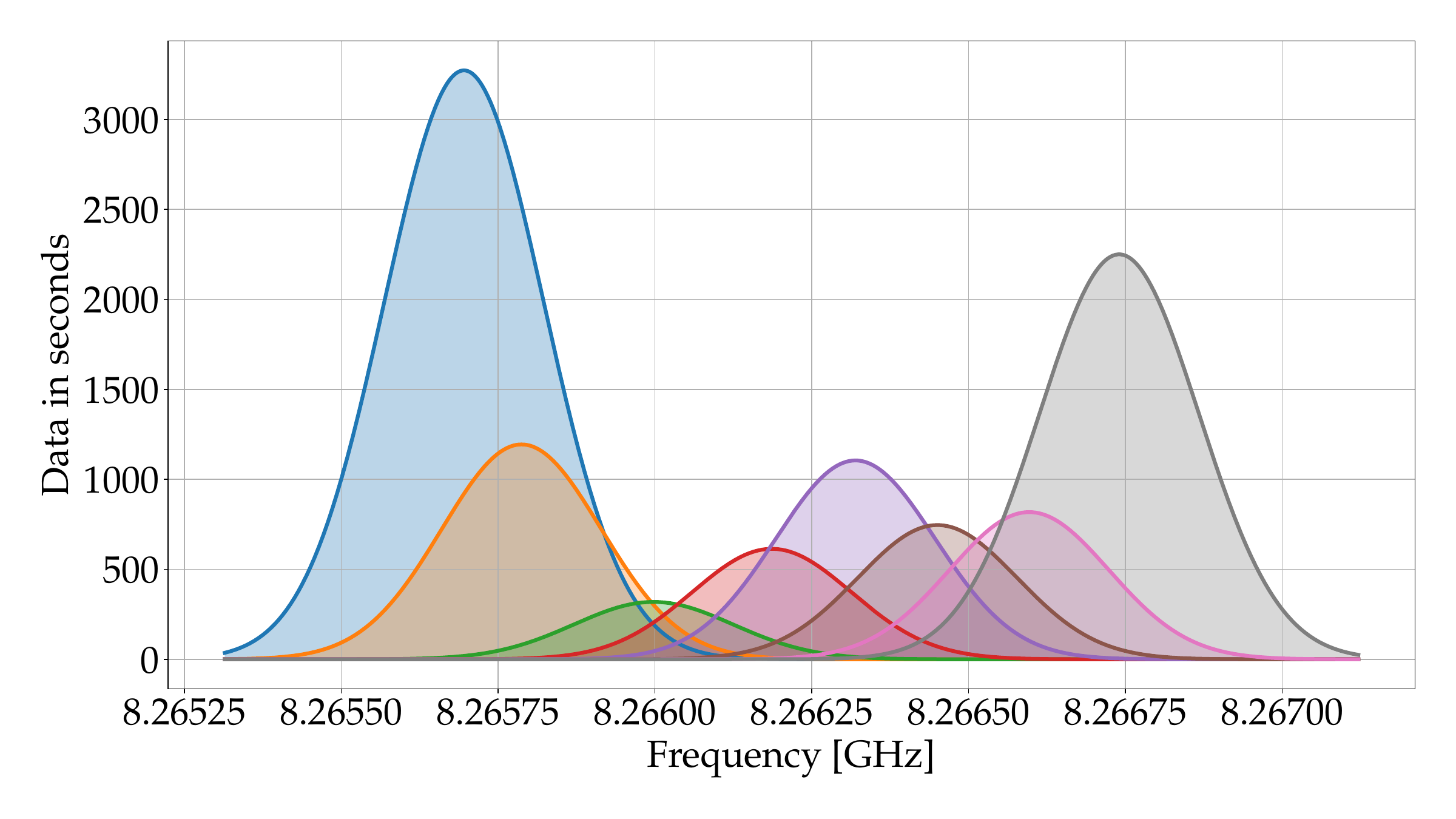}
    \caption{Amount of collected data in seconds per frequency interval. For each shaded area the cavity is tuned to the centre frequency of the plotted resonance curve, where the width is taken from the measurement. The amplitude of each curve corresponds to the measurement time at that frequency. }
    \label{fig:axion_dataOverview}
\end{figure}

The cryostat was operated at 2.0\,K with the pressure ranging from 30\,mbar to 90\,mbar. As described in \cite{SupaxCDR}, varying the pressure enables tuning of the resonance frequency, which allowed scanning a frequency range of approximately $\SI{1.4}{MHz}$. The effective measurement time as a function of the set frequency is shown in Fig.\,\ref{fig:axion_dataOverview}.

\subsection{Constructing the grand unified spectrum}
The construction of the grand unified spectrum (GUS) is briefly explained in the following, while it is described in detail in \CDR. First, the recorded spectra for a given measurement frequency interval, as illustrated in Fig. \ref{fig:axion_dataOverview}, are integrated. 
The resulting \textit{raw} spectra $\delta_{ik}^r$\footnote{The superscript in $\delta_{ik}^r$ refers to the processing stage: $r$ = raw, $f$ = filtered, $s$ = rescaled by resonance curve, $g$ = GUS.}, where $i$ denotes the index of the spectrum and $k$ denotes the frequency bin within each spectrum, are processed with a Savitzky–Golay (SG) filter \cite{Savitzky:2002vxy} to remove the cavity's resonance and electronic residual structures 
\begin{align}
    \delta_{ik}^f = \frac{\delta_{ik}^r}{\delta_{ik}^{SG}} - 1 \label{eq:def_SG_filtered_spectra}.
\end{align}
The \textit{filtered} spectra $\delta_{ik}^f$ are rescaled by dividing them by the corresponding cavity Lorentzian 
\begin{equation}    \label{eq:scaling_factor_rades_gus}
       L_{ik}(f) = \frac{1}{1+4Q_L^2(f/f_0^i - 1)^2}
\end{equation}
normalized to 1 at the resonance frequency where every spectrum $i$ has its own resonance frequency $f_0^i$. 
For these re\textit{scaled} spectra $\delta_{ik}^s$, the noise of every bin $\sigma_{ik}^s$ is scaled as 
\begin{gather}
    \delta_{ik}^s = \frac{\delta_{ik}^f}{L_{ik}} \label{eq:rescaled_spectra} \quad \text{and} \quad \sigma_{ik}^s = \frac{\sigma_i^f}{L_{ik}}\,,
\end{gather} 
where $\sigma_i^f$ is the standard deviation of the $i$-th spectrum.

\begin{figure}[ht]
    \centering
    \begin{subfigure}[b]{0.49\textwidth}
        \includegraphics[width=\textwidth]{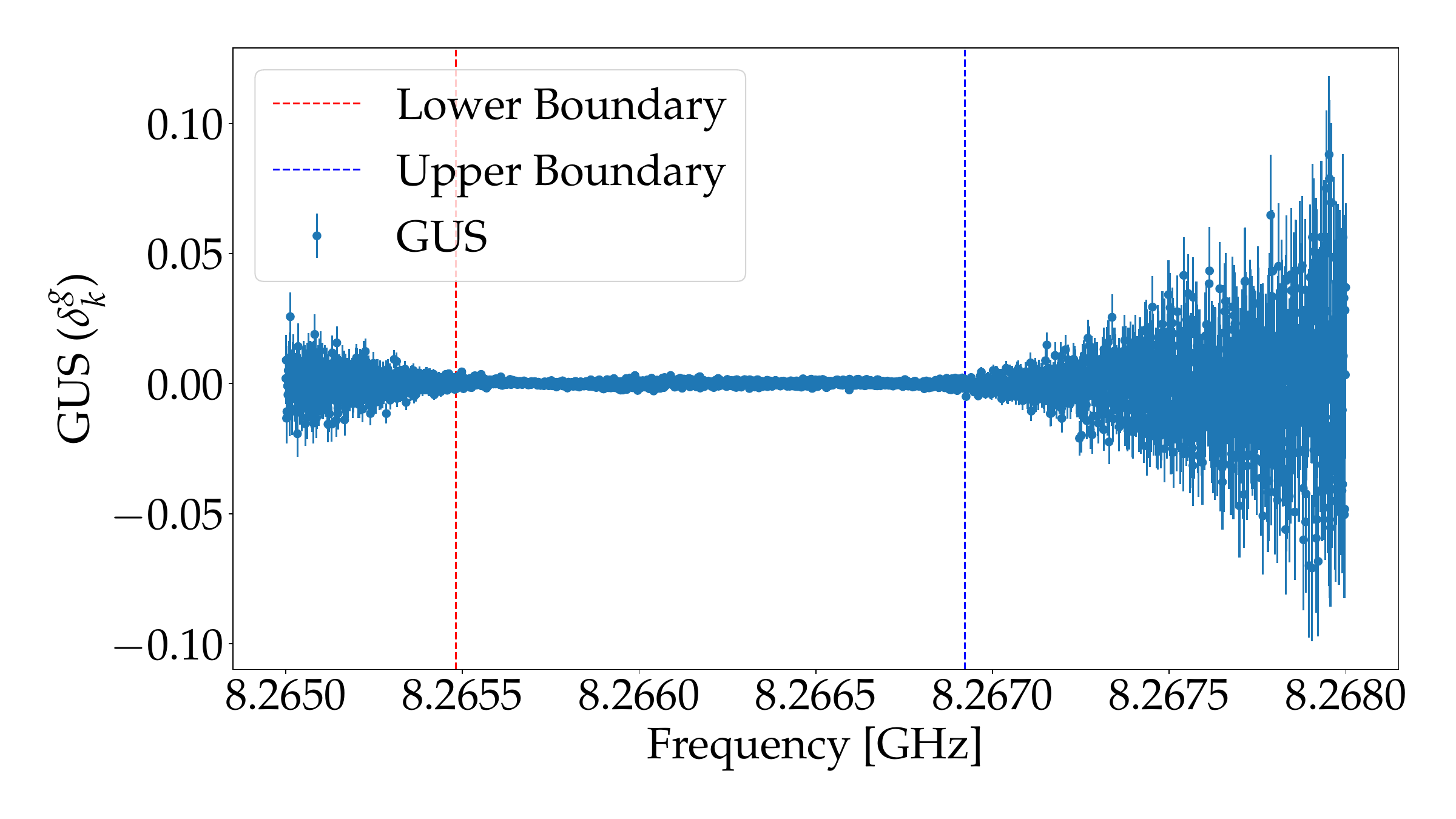}
    \end{subfigure}
    \hfill
    \begin{subfigure}[b]{0.49\textwidth}
        \includegraphics[width=\textwidth]{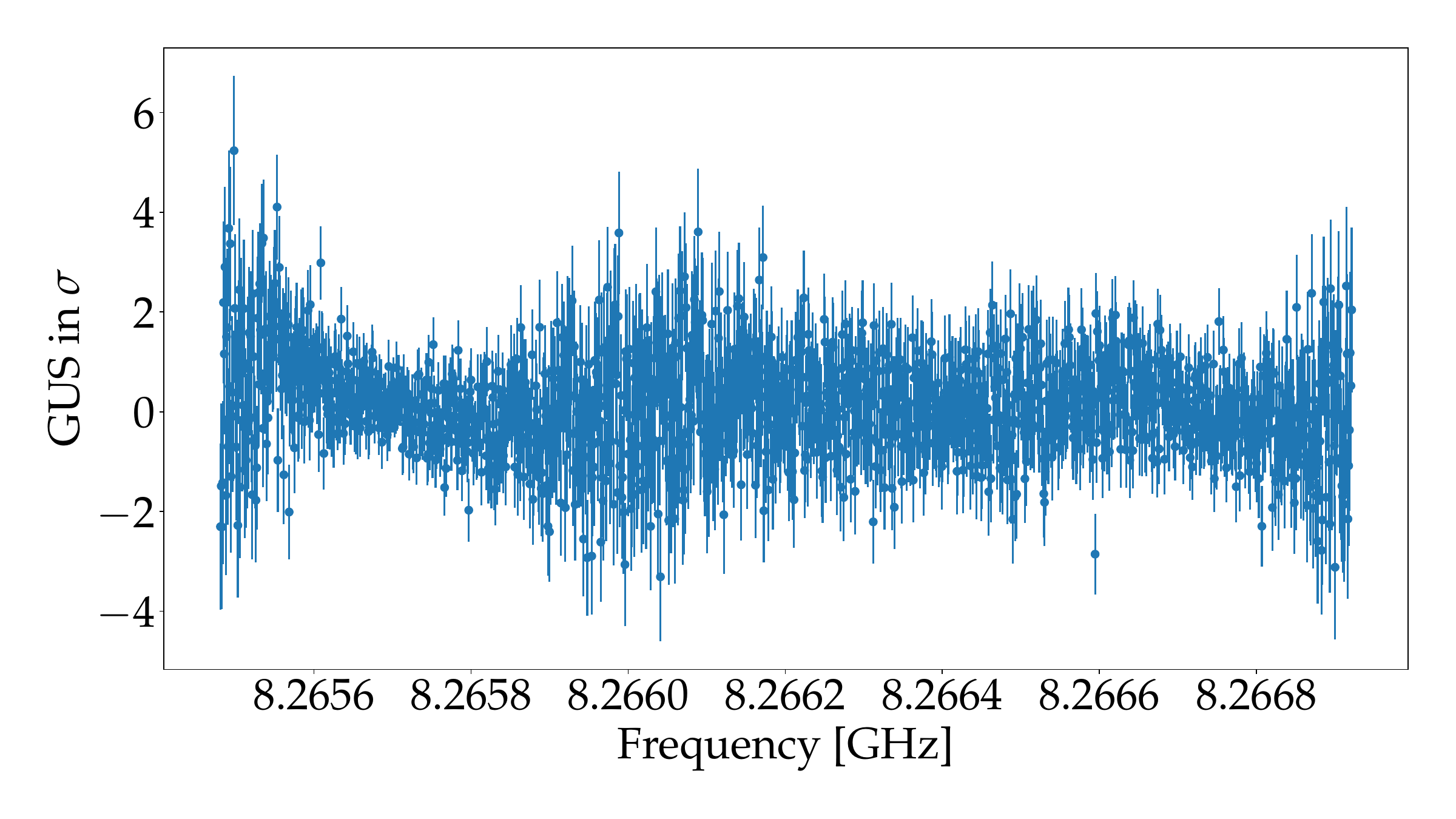}
    \end{subfigure}
    \caption{Grand unified spectrum of the analysed data. Left: Absolute values of the grand unified spectrum over the entire frequency range, where most sensitive range is between the vertical dashed lines. Right: Zoom into the most sensitive range showing the normalized values in units of $\sigma_k^g$. Small absolute values of GUS indicate high sensitivity.}
    \label{fig:axion_GUS}
\end{figure}

Finally, the GUS $\delta_k^g$ with uncertainty $\sigma_k^g$ is built by adding all spectra weighted by their inverse variance
\begin{gather}
    \delta_k^g = \sum_i \delta_{ik}^s \cdot \omega_{ik} \, \text{;} \quad
    \omega_{ik} = \frac{(\sigma_{ik}^s)^{-2}}{\sum_i (\sigma_{ik}^s)^{-2}}\,, \\
    \sigma_k^g = \sqrt{\sum_i (\sigma_{ik}^s)^2 \cdot \omega_{ik}^2} \label{eq:GUS_err_calc}\,.
\end{gather}
The resulting GUS over the whole frequency range, as well as a zoom into the most sensitive range, is shown in Fig.\,\ref{fig:axion_GUS}

The GUS is now considered only in the sensitive area of the experiment between 8.2655\,GHz and 8.2669\,GHz. 
In the absence of a dark matter signal the GUS should be nothing but white noise. 
Thus we can compare the expected (indicated by daggered variables) power fluctuation $\sigma_k^\dag = \frac{1}{\sqrt{\Delta \nu \cdot \tau}}$ to the measured one $\sigma_k^g$ where $\Delta \nu$ is the bin width and $\tau$ the total integration time. 
Since in the construction of the GUS we rescale $\sigma_k^g$ in eq. \eqref{eq:rescaled_spectra} we need to rescale $\sigma_k^\dag$ in the same fashion to compare them $(\sigma_k^s)^\dag = \sigma_k^\dag/\bar{L}_{k}$.

\begin{figure}[ht]
    \centering
    \includegraphics[width=0.5\textwidth]{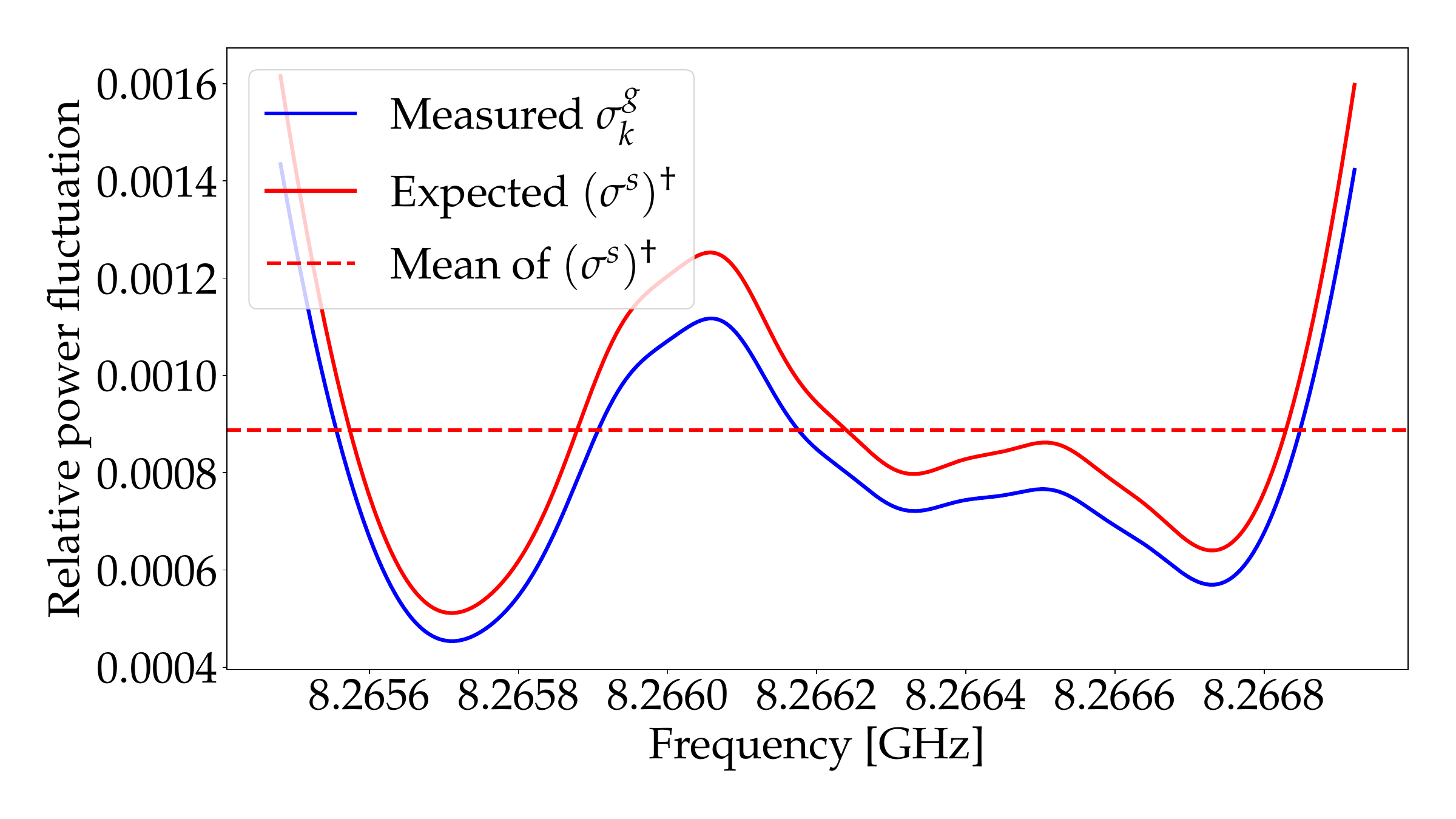}
    \caption{Rescaled expected [$(\sigma^s)^\dag$] and measured [$\sigma_k^g$] noise power fluctuations agree apart from an offset that is originating from the maximum likelihood weighting of the measured GUS giving an edge over the simple integration time weighting of the expected noise.}
    \label{fig:comp_theo_exp_noise}
\end{figure}

\subsection{Limit setting}
The limit-setting procedure follows the approach described in \CDR. Under these assumptions, the axion-induced signal is described by the lineshape \cite{Axion_lineshape1990}
\begin{gather}
    f(\nu)=\frac{2}{\sqrt{\pi}}\left(\sqrt{\frac{3}{2}} \frac{1}{r} \frac{1}{\nu_a\left\langle\beta^2\right\rangle}\right) \sinh \left(3 r \sqrt{\frac{2\left(\nu-\nu_a\right)}{\nu_a\left\langle\beta^2\right\rangle}}\right) \exp \left(-\frac{3\left(\nu-\nu_a\right)}{\nu_a\left\langle\beta^2\right\rangle}-\frac{3 r^2}{2}\right)\,.
    \label{eq:virialized_mass_peak}
\end{gather}
Here, $\nu_a$ denotes the frequency corresponding to the rest mass of the axion, and $\langle \beta^2 \rangle = \langle v^2 \rangle / c^2$ is the second moment of the Maxwell–Boltzmann velocity distribution, defined as $\langle v^2 \rangle = \frac{3}{2} v_c^2$. 
The local galactic circular velocity is taken to be $v_c = \SI{220}{km/s}$, which sets the characteristic velocity dispersion of virilized axion dark matter in the Solar neighbourhood.

The distance from the galactic centre is given by $r = v_c^2 / \langle v^2 \rangle \approx \sqrt{2/3}$ \cite{PhysRevD.42.3572}.
The continuous lineshape is then discretized as
\begin{gather}
    D_q = \int_{\nu_a + (q-1) \Delta \nu}^{\nu_a + q \Delta \nu} f(\nu) \, \dd \nu \label{eq:discretized_mass_peak}
\end{gather}
and subsequently processed with a Savitzky–Golay (SG) filter using the same parameters as those applied during the construction of the GUS. This procedure accounts for the impact of the SG filter on the lineshape of a potential dark matter signal. The resulting SG-filtered signal shape, denoted by $D_q^S$, can then be searched for in the GUS following the same methodology described in \CDR.
Assuming that $\delta_k^g$ is Gaussian distributed, the likelihood function for each bin $k$ converges to a $\chi^2$ distribution of $\delta_k^g$, with the signal model given by $y(k, A) = A \cdot D_q^S$. The likelihood function is therefore written as
\begin{gather}
    L = e^{-\frac{1}{2} S(\delta_k^g, A)}, \quad \text{with} \quad
    S(\delta_k^g, A) = \underset{k}{\sum} \frac{\left[\delta_k^g - y(k,A)\right]^2}{\sigma_k^2}\,.
\end{gather}
$S(\delta_k^g, A)$ follows a $\chi^2$ distribution. The upper limit on the signal strength $A_{\text{UL}}$ is found by solving
\begin{gather}
    1 - \alpha = \frac{\int_{0}^{A_\text{UL}} e^{-\frac{1}{2}S(\delta_k^g, A)} \, \dd A}{\int_{0}^{\infty} e^{-\frac{1}{2}S(\delta_k^g, A)} \, \dd A} \,. \label{eq:A_UL_calc}
\end{gather}
The upper limit $A_{\text{UL}}$ is determined for a significance level of $\alpha = 0.05$, corresponding to a 95\% confidence level. 
Interpreting $A_{\text{UL}}$ as a measure of the excess signal-to-noise ratio, the maximum signal power $P_S$ that can be excluded in the measured data is determined for each bin $k$ as $P_S^k = A_\text{UL}^k \cdot P_N^k$, where the noise power is given by $P_N = k_B , T_{\text{sys}} , \Delta \nu$. Using $P_S^k$, the corresponding axion–photon coupling is extracted using Eq.\,\ref{eq:PS_axions}. The experimental parameters entering this extraction are summarized in Table~\ref{tab:experimental_parameters}.

\begin{figure}[ht]
    \centering
    \includegraphics[width=0.7\linewidth]{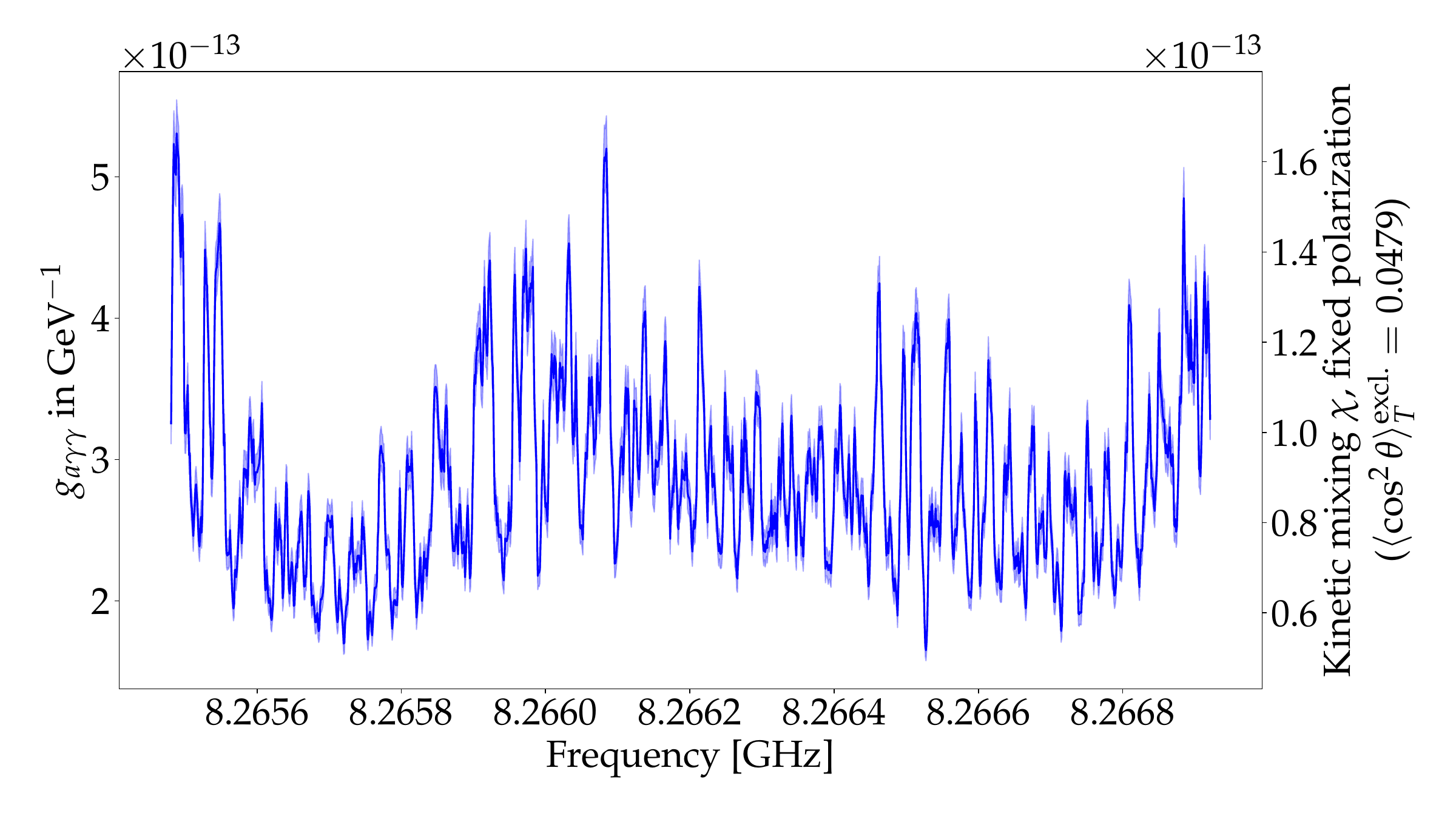}
    \caption{Measured limit on axion coupling $\abs{g_{a\gamma\gamma}}$ (left axis) and kinetic mixing parameter $\chi$ assuming fixed polarization (right axis) in dependence of the frequency. The limit on $\chi$ assuming random polarization is $6.96$ times lower. In the figure a constant frequency and therefore fixed recast factor between $\abs{g_{a\gamma\gamma}}$ and $\chi$ is assumed, yielding a negligible systematic shift for low and high frequencies of $0.01\%$ on $\chi$.}
    \label{fig:axion_limit_plot}
\end{figure}

The noise power per bin, $P_N$, is determined by the system noise temperature $T_{\text{sys}}$ and the bin width $\Delta \nu$ according to Eq.~\ref{eq:snr}. For each bin $k$, the value of $A_{\text{UL}}$ is obtained numerically. The resulting exclusion limits on the axion–photon coupling are shown in Fig.~\ref{fig:axion_limit_plot}. These limits range from $1.6 \times 10^{-13}\si{GeV}^{-1}$ to $5.3 \times 10^{-13}\si{GeV}^{-1}$ over a frequency band of \SI{1.44}{MHz} centred at $f_0 = \SI{8.2662}{GHz}$.

%

\begin{table}[ht]
    \centering
    \begin{tabular}{c | c | c}
        \toprule
        Symbol & Quantity & Value \\
        \midrule
        $\rho_\text{DM}$ & local DM density & $0.45~\si{GeV\per\cubic{\centi\meter}}$ \\
        $V$ & effective volume & $(102600 \pm 310) \, \si{\cubic\milli\meter}$ \\
        $C$ & Overlap of $TM_{010}$ mode with B-field & 0.549 \\
        $\langle \cos^2 \theta \rangle_T$ & random polarization exclusion factor & $1/3$ \\
        $\excl$ & fixed polarization exclusion factor & $0.0479$ \\
        $Q_L$ & loaded quality factor & $27620 \pm 650$  \\
        $\beta$ & coupling factor & $0.87 \pm 0.06$  \\
        $\eta$ & cable signal attenuation & $0.85 \pm 0.05$  \\
        $T_\text{sys}$ & system noise temperature & $(10 \pm 0.5) \, \si{K}$  \\
        $\Delta \nu$ & bin width & \SI{960}{Hz}  \\
        $\tau$ & total integration time & \SI{10318}{s}  \\
        $B$ & external magnetic field & $(12.0 \pm 0.1) \, \si{T}$ \\
        \bottomrule
    \end{tabular}
    \caption{Summary of experimental parameters and uncertainties.}
    \label{tab:experimental_parameters}
\end{table}


Assuming no signal contribution from axions, the determined limits on the signal power $A_{\text{UL}}$ can be directly used to set limits on the kinetic mixing parameter of the dark photon by adjusting the signal model accordingly.
The coupling limit found for axions $\abs{g_{a\gamma\gamma}}$ is then recalculated into a limit on the dark photon kinetic mixing parameter $\chi$ following \cite{SupaxCDR} as 
\begin{gather}
    \chi^2 = \abs{g_{a\gamma\gamma}}^2 \frac{B^2 \varepsilon_0 \hbar^3 c^5}{m_\text{DM}^2 \excl}\,. \label{eq:limit_recast}
\end{gather}
In the fixed-polarization scenario, the expected overlap factor of the dark photon field with the cavity axis $\excl$ is calculated for a finite data acquisition time T as described in \cite{Caputo_2021} for the chosen confidence level of $95\%$. In the random polarization scenario this factor becomes $1/3$. All other parameters are given in Table \ref{tab:experimental_parameters}.
The recast limits on the DP kinetic mixing parameter are presented on the right side of Fig.\,\ref{fig:axion_limit_plot} and range between $\chi < 2\cdot 10^{-14}$ to $\chi < 6.3\cdot 10^{-14}$ for the random-polarization scenario and $\chi < 5.2\cdot 10^{-14}$ to $\chi < 1.6\cdot 10^{-13}$ for the fixed-polarization scenario.

\begin{figure}[ht]
    \centering
    \includegraphics[width=0.7\linewidth]{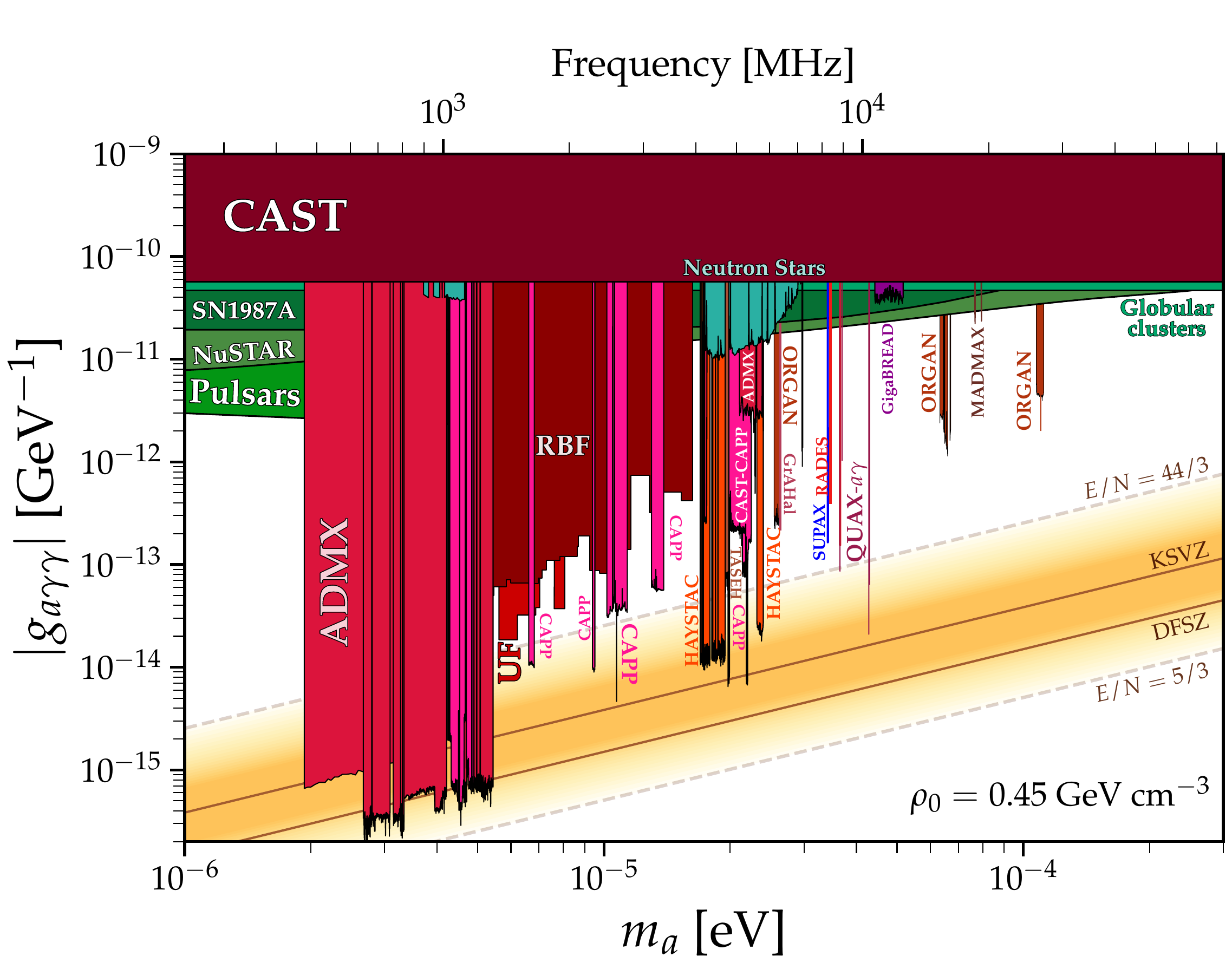}
    \caption{The measured axion-photon limit displayed in blue at $m_a = 34\,\mu\text{eV}$ in comparison to other results taken from \cite{AxionLimits}.}
    \label{fig:axion_UpperLimits}
\end{figure}

\begin{figure}[ht]
    \centering
    \includegraphics[width=0.85\linewidth]{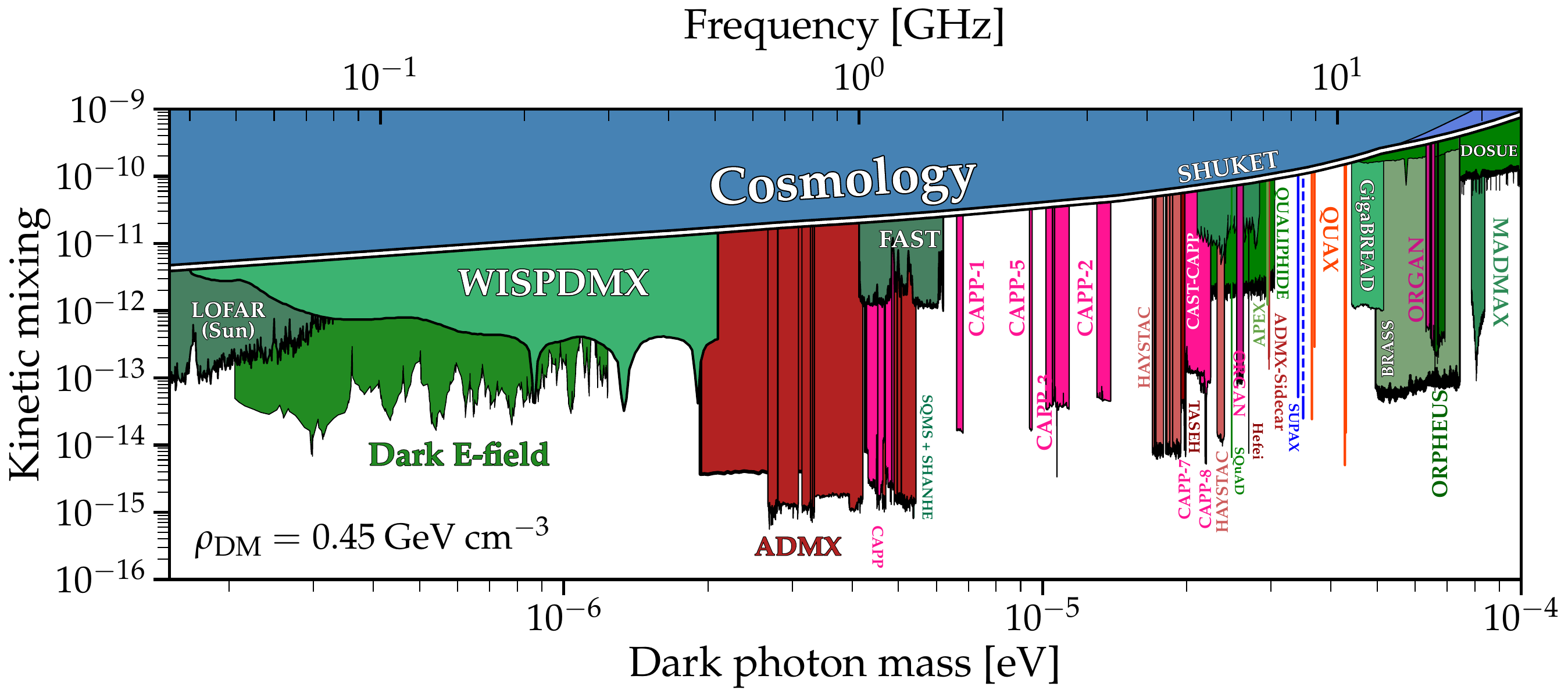}
    \caption{The measured limit on the dark photon kinetic mixing parameter in the fixed-polarization scenario, shown in blue at $m = 34\,\muev$, in comparison to other results taken from \cite{AxionLimits}, including the dedicated \Supax dark photon results (dashed line) from \cite{SupaxCDR}.}
    \label{fig:DP_UpperLimits}
\end{figure}

The results presented in this work are compared with existing experimental constraints in Fig.\,\ref{fig:axion_UpperLimits} for the axion–photon coupling and in Fig.\,\ref{fig:DP_UpperLimits} for the fixed-polarization dark photon kinetic mixing parameter. Despite the limited data-taking period, competitive exclusion limits are obtained from the initial \SI{3}{h} measurement run of the \Supax prototype experiment. These results were achieved using a novel tuning mechanism (tuning the helium gas pressure) that enabled efficient coverage of a \SI{1.4}{MHz}-wide frequency range, demonstrating the viability and discovery potential of the experimental concept.


\newpage
\newpage
\section{Conclusion and Outlook}
\label{sec:outlook}

The SUPerconduction AXion search experiment (\Supax) is a planned haloscope experiment aimed at detecting axion-like particles (ALPs) at the University of Bonn, which are compelling candidates for dark matter and offer a potential solution to the strong-CP problem. Targeting the mass range between $8\,\muev$ and 30\,$\mu $eV, \Supax will utilize superconducting cavities in a strong magnetic field with advanced tuning mechanisms to achieve high sensitivity to ALP-induced signals. 
As a preparatory step, a prototype experiment was constructed and operated, employing a copper cavity with a quality factor of $5\cdot10^4$, cooled to a temperature of 2\,K, and placed within a 12.1\,T magnetic field. 
A narrow mass range of 0.006\,$\mu$eV (1.4\,MHz) width around $34\,\muev$ was probed by tuning the cavity resonance frequency via precise changes of the gas pressure. Axion--photon couplings larger than $\abs{g_{a\gamma\gamma}} > 1.6\cdot10^{-13}$ are excluded at 95\% confidence level under the stated assumptions. With this result, approximately three years after the idea of the \Supax experiment, we can report on the successful design and construction of an RF-cavity, the entire data-acquisition system, and data analysis. 

%
%
With the gained experience, a new Sikivie-type haloscope is being presently developed. Two setups will be constructed, operating at a temperature of 10\,mK and providing a 12\,T magnetic field while utilising near quantum limited readout for the initial phase. One setup will be constructed at the university of Mainz and the second one at the university of Bonn, both to be commissioned by the end of 2026. 
While the haloscopes will be used to search for axions, as detailed in \cite{SupaxCDR}, the main focus will become the search for high-frequency gravitational waves within the sensor network, GravNet \cite{Schmieden:2023fzn}.

\section*{Acknowledgement}
The work of Tim Schneemann was funded by the DFG graduate school 2796 ``particle detectors". This work was supported by the DFG Project ID 390831469: EXC 2118 (PRISMA++ Cluster of Excellence), by the COST Action within the project COSMIC WISPers (Grant No. CA21106), and by ERC grant ERC-2024-SYG 101167211, as well as the preparatory research funds of the ColorMeetsFlavor Cluster of Excellence.

\bibliographystyle{apsrev4-1} 
\bibliography{./Bibliography}
	
\end{document}